\documentclass[epjST]{svjour}
\usepackage{graphics}
\usepackage{amssymb} 

\usepackage{color}

\begin{document}

\journalname{EPJ -- Special Topics (2011) /}

\issuename{\small Discussion \& Debate: \emph{From black swans to dragon kings. Is there life beyond power laws?}}

\title{Black swans or dragon kings?
A simple test for deviations from the power law}

\author{Joanna Janczura\inst{1}\fnmsep\thanks{\email{joanna.janczura@pwr.wroc.pl}} 
\and 
Rafa{\l} Weron\inst{2}\fnmsep\thanks{\email{rafal.weron@pwr.wroc.pl}}}

\institute{Hugo Steinhaus Center for Stochastic Methods, Wroc{\l}aw University of Technology, Poland 
\and 
Institute of Organization and Management, Wroc{\l}aw University of Technology, Poland}

\abstract{We develop a simple test for deviations from power law tails, which is based on the asymptotic properties of the empirical distribution function. We use this test to answer the question whether great natural disasters, financial crashes or electricity price spikes should be classified as dragon kings or `only' as black swans. 
} 
\maketitle

\section{Introduction}
\label{sec:intro}

In a recent article Sornette \cite{sor:09} presents a number of data sets exhibiting power laws with extreme outliers. He calls these events `dragon kings' and says that they are the result of positive feedback mechanisms that make them much larger than their peers. Being outliers, these dragon kings are unaccounted for by power laws, for which Taleb \cite{tal:07} coined the nowadays popular term `black swans'. According to Sornette, extreme events are significantly more likely to happen than power laws suggest. Are dragon kings then `to blame' for experiencing two or three once-in-a-millennium floods or financial crises in the last couple of decades? What is a dragon king anyway? And how do we know that a given observation is a dragon king and not simply a `normal' deviation to be expected of any random sample? Definitely these issues are intriguing and worth an investigation.

Having the latter question in focus, in Section \ref{sec:test} we develop a simple test for deviations from power law tails. Actually, from the tails of any distribution. The test is based on the asymptotic properties of the empirical distribution function (edf) and -- as we show in Section \ref{sec:examples} -- is a universal tool that can be used in many practical situations.

\section{Testing for dragon kings}
\label{sec:test}

The simple test we propose here is based on the asymptotic properties of the empirical distribution function (edf): 
\begin{equation}
F_n(x)=\frac{1}{n}\sum_{i=1}^n\mathbb{I}_{\{x_i<x\}},
\end{equation}
which is defined for a sample of observations $(x_1,x_2,...,x_n)$, with $\mathbb{I}$ being the indicator function. 
By the Central Limit Theorem, $F_n(x)$ is asymptotically normally distributed \cite{bic:dok:01}. Precisely,
\begin{equation}\label{eqn:CLT:edf}
\frac{\sqrt{n}[F_n(x)-F(x)]}{\sqrt{F(x)[1-F(x)]}} \stackrel{d}{\rightarrow} N(0,1),
\end{equation}
where $F(x)$ is the true cumulative distribution function (cdf) and $\stackrel{d}{\rightarrow}$ denotes convergence in distribution. 

Using property (\ref{eqn:CLT:edf}), we can construct confidence intervals (CI) for the edf and an arbitrarily chosen confidence level $(1-\alpha)$. Note, that these are pointwise intervals, meaning that for each specified value of $x$, we are $(1-\alpha)\times 100\%$ confident of observing edf($x$) within those limits. This is not the same as constructing the so-called confidence bands which guarantee, with a given confidence level, that the edf falls within the band for all $x$'s (in some interval); these bands are wider than the curves one obtains by using pointwise CI \cite{bor:98,kle:moe:03}.

Letting $z_{\frac{\alpha}{2}}$ and $z_{1-\frac{\alpha}{2}}$ denote the $(\frac{\alpha}{2})$ and $(1-\frac{\alpha}{2})$-quantiles of the standard normal distribution, respectively, we have:
\begin{equation}\label{eqn:edf:approx}
P\left(z_{\frac{\alpha}{2}}< \frac{\sqrt{n}[F_n(x)-F(x)]}{\sqrt{F(x)[1-F(x)]}} < z_{1-\frac{\alpha}{2}}\right)\approx 1-\alpha, 
\end{equation}
provided that $n$ is large enough. Note, that since the standard Gaussian law $N(0,1)$ is symmetric around 0, $z_{1-\frac{\alpha}{2}} = - z_{\frac{\alpha}{2}}$. Formula (\ref{eqn:edf:approx}) implies that: 
\begin{equation}\label{eqn:edf:lt}
P\left(F(x)+\sqrt{\frac{F(x)[1-F(x)]}{n}}z_{\frac{\alpha}{2}}  < F_n(x) < F(x)+\sqrt{\frac{F(x)[1-F(x)]}{n}}z_{1-\frac{\alpha}{2}}  \right)\approx 1-\alpha.
\end{equation}
Analogously, for the right tail we have:
\begin{eqnarray}
P\left(1-F(x)+\sqrt{\frac{F(x)[1-F(x)]}{n}}z_{\frac{\alpha}{2}}  < 1-F_n(x) < \right. \qquad\qquad \nonumber\\
\left. < 1-F(x)+\sqrt{\frac{F(x)[1-F(x)]}{n}}z_{1-\frac{\alpha}{2}}  \right)\approx 1-\alpha.\label{eqn:edf:rt}
\end{eqnarray}

At this point assume that the true distribution $F$ has power law tails, i.e.\ $F(x)\approx b_1x^{p_1}$, for $x\rightarrow -\infty$, and $1-F(x)\approx b_2x^{p_2}$, for $x\rightarrow \infty$. 
From (\ref{eqn:edf:lt}), the left tail of the edf should lie in the interval 
\begin{equation}\label{eqn:left_tail:int}
\left(b_1x^{p_1}+\sqrt{\frac{b_1x^{p_1}(1-b_1x^{p_1})}{n}}z_{\frac{\alpha}{2}},\ b_1x^{p_1}+\sqrt{\frac{b_1x^{p_1}(1-b_1x^{p_1})}{n}}z_{1-\frac{\alpha}{2}}\right)
\end{equation}
with probability $1-\alpha$. 
Similarly, from (\ref{eqn:edf:rt}) the right tail should lie in the interval 
\begin{equation}\label{eqn:right_tail:int}
\left(b_2x^{p_2}+\sqrt{\frac{b_2x^{p_2}(1-b_2x^{p_2})}{n}}z_{\frac{\alpha}{2}},\ b_2x^{p_2}+\sqrt{\frac{b_2x^{p_2}(1-b_2x^{p_2})}{n}}z_{1-\frac{\alpha}{2}}\right)
\end{equation}
with probability $1-\alpha$. 

Now, it suffices to fit a power law to the left or right tail of the edf built from the analyzed sample and plot the respective intervals (\ref{eqn:left_tail:int}) or (\ref{eqn:right_tail:int}). Observations lying outside the curves spanned by the CI are likely to be (i.e.\ with probability $1-\alpha$) dragon kings. Note, that the presented approach is general. The true distribution can be arbitrary, say, stretched exponential (also known as Weibull). Only then the intervals (\ref{eqn:left_tail:int})-(\ref{eqn:right_tail:int}) would be computed from relations (\ref{eqn:edf:lt})-(\ref{eqn:edf:rt}) using the stretched exponential cdf.

\section{Empirical examples}
\label{sec:examples}

\subsection{Simulated data}

We start the empirical analysis with a simulation study to check the effectiveness of the test for dragon kings. The results summarized in Table \ref{tab:sim:power} concern random samples from two heavy-tailed distributions with power law decay in the tail(s):
\begin{itemize}
\item 
Cauchy with cdf
$
F(x;\mu,\sigma)=\frac{1}{\pi}\arctan\left( \frac{x-\mu}{\sigma} \right)+\frac{1}{2},
$
\item 
Pareto with cdf
$
F(x;\lambda,\alpha)=1-\lambda^\alpha (x+\lambda)^{-\alpha},
$
\end{itemize}
and two lighter tailed laws:
\begin{itemize}
\item 
symmetric Hyperbolic with the cdf obtained by numerically integrating the probability density function (pdf)
$
f(x;\alpha,\delta)=\exp(-\alpha\sqrt{\delta^2+x^2}+\beta x) / \{2\delta K_1(\delta\alpha)\},
$
where $K_1$ is the modified Bessel function of the third kind,
\item 
stretched exponential (or Weibull) with cdf
$
F(x;\beta,\tau)=1-e^{-\beta x^{\tau}}.
$
\end{itemize}
For simulation and estimation issues, as well as, sample applications of these distributions in finance and insurance see e.g.\ \cite{ciz:hae:wer:11,klu:pan:wil:08,pao:07}.

\begin{table}
\caption{Percentage of pointwise outliers with respect to the CI given by formula (\ref{eqn:edf:rt}) and $F(x)$ being a power law fitted to the 10\%-1\% or 25\%-2.5\% largest observations. For Cauchy and Pareto distributions, outliers with respect to the true power law (implied by the cdf) are also provided. The number of simulated samples is equal to $10^4$.}
\label{tab:sim:power}
\centering\small      
\begin{tabular}{ccrrrrrr}
\hline\noalign{\smallskip}
CI & Sample & \multicolumn{2}{r}{Cauchy(0,1)} &  \multicolumn{2}{r}{Pareto(2,1)} & Hyp(2,1) & Weib(1,$\frac{1}{2}$)\\
& size & True & Fitted & True & Fitted & Fitted & Fitted \\
\noalign{\smallskip}\hline\noalign{\smallskip}
&&\multicolumn{6}{c}{\it Power law fitted to 10\%-1\% largest observations} \\ \noalign{\smallskip}
90\% & 1000 & 9.6\% & 4.2\% & 9.1\% & 4.3\% & 29.2\% & 20.4\% \\
95\% & 1000 & 4.8\% & 2.1\% & 4.1\% & 1.7\% & 10.1\%  & 6.1\%\\
99\% & 1000 & 1.1\% & 0.6\% & 0.9\% & 0.4\% & 0.1\% & $<$0.1\% \\
\noalign{\smallskip}\hline\noalign{\smallskip}
&&\multicolumn{6}{c}{\it Power law fitted to 25\%-2.5\% largest observations} \\ \noalign{\smallskip}
90\% & 1000 & " &10.7\%  & " &  11.2\%  & 99.1\% & 90.9\% \\
95\% & 1000 & " & 4.4\%  & " &  4.7\%   & 99.8\% & 78.1\% \\
99\% & 1000 & " & 0.7\%  & " &  0.5\%   & 74.3\% & 35.0\% \\
\noalign{\smallskip}\hline\noalign{\smallskip}
&&\multicolumn{6}{c}{\it Power law fitted to 10\%-1\% largest observations} \\ \noalign{\smallskip}
90\% & 5000 & 10.0\% & 9.9\%& 9.8\% & 11.3\%& 99.8\%&  98.9\%\\
95\% & 5000 &  4.6\% & 4.3\%& 4.7\% &  5.1\%& 99.5\%&  96.5\%\\
99\% & 5000 &  1.0\% & 1.2\%& 1.0\% &  0.6\%& 93.3\%&  78.9\%\\
\noalign{\smallskip}\hline
\end{tabular}
\end{table}

\begin{table}
\caption{Percentage of pointwise outliers with respect to the CI given by formula (\ref{eqn:edf:rt}) and $F(x)$ being a stretched exponential (or Weibull) law fitted to the 10\%-1\% or 25\%-2.5\% largest observations. For the Weibull distribution, outliers with respect to the tail of the true cdf are also provided. The number of simulated samples is equal to $10^4$.}
\label{tab:sim:weib}
\centering\small      
\begin{tabular}{ccrrrrr}
\hline\noalign{\smallskip}
CI & Sample & Cauchy(0,1) &  Pareto(2,1) & Hyp(2,1) & \multicolumn{2}{r}{Weib(1,$\frac{1}{2}$)}\\
& size & Fitted & Fitted & Fitted & True & Fitted \\
\noalign{\smallskip}\hline\noalign{\smallskip}
&&\multicolumn{5}{c}{\it Weibull tail fitted to 10\%-1\% largest observations} \\ \noalign{\smallskip}
90\% & 1000& 62.1\% &58.3\% &11.5\% & 9.6\%& 10.6\% \\
95\% & 1000& 56.4\% &52.4\% & 4.9\% & 4.8\%&  5.8\% \\
99\% & 1000& 44.8\% &41.3\% & 0.6\% & 1.1\%&  2.4\% \\
\noalign{\smallskip}\hline\noalign{\smallskip}
&&\multicolumn{5}{c}{\it Weibull tail fitted to 25\%-2.5\% largest observations} \\ \noalign{\smallskip}
90\% & 1000&96.7\% &97.5\%& 63.2\% &"&14.0\% \\
95\% & 1000&95.3\% &96.6\%& 47.7\% &"& 8.0\% \\
99\% & 1000&91.8\% &94.1\%& 19.0\% &"& 2.6\%  \\
\noalign{\smallskip}\hline
\end{tabular}
\end{table}

The validation procedure is the following. We calculate the edf for $10^4$ simulated samples from each of the four distributions and two sample sizes. Next, using least squares regression we fit a power law to the 10\%-1\% (or 25\%-2.5\%) largest observations. The upper 1\% (or 2.5\%) of values are not used for calibration due to the sensitivity of the least squares fit to outlying observations. Then, we compute the pointwise CI given by formula (\ref{eqn:right_tail:int}). Finally, we check whether for a particular value of $x$ the edf lies within the CI. Motivated by the fact that dragon kings are to be expected in the very tails of the distribution, we arbitrarily set $x$ to be the fourth largest observation, i.e.\ $x_{(4)}$. However, qualitatively identical results are obtained for $x_{(8)}$ and $x_{(12)}$.

\newpage

The percentages of pointwise outliers at $x=x_{(4)}$ with respect to the computed CI are given in Table \ref{tab:sim:power}. As we can observe, the results are highly dependent on the subsample used to calibrate the power law. For the larger sample size (5000 observations), the 10\%-1\% range leads to accurate coverage rates for the Cauchy and Pareto laws, while for the lighter tailed distributions $x_{(4)}$ is classified as an outlier nearly in all samples. For the smaller sample size (1000 observations), the power law fitted to the 10\%-1\% range yields overly conservative CI for the Cauchy and Pareto laws -- the rejection rates are roughly twice lower. In such a case, the 25\%-2.5\% range has to be used to obtain a reliable coverage by the CI. However, the percentages of pointwise outliers with respect to the true power law (implied by the parameters of cdf) are accurate. The latter indicates that the dragon king test works well -- even for small sample sizes -- provided that the power law is correctly estimated. Otherwise it may lead to a lower rejection rate and, hence, a lower number of observations classified as outliers. 

As we have said previously, the test of Section \ref{sec:test} is universal in the sense that the true distribution can be arbitrary. We thus repeat the simulation study with another popular heavy-tailed, but not power-law tailed, distribution -- the stretched exponential. Note, that in actuarial sciences and statistics it is more commonly known as the Weibull law \cite{ciz:hae:wer:11,klu:pan:wil:08}. Now the intervals (\ref{eqn:left_tail:int})-(\ref{eqn:right_tail:int}) are computed from relations (\ref{eqn:edf:lt})-(\ref{eqn:edf:rt}) using the Weibull cdf. To save space, the results summarized in Table \ref{tab:sim:weib} concern only the smaller sample size (1000 obs.), which better reflects the sizes of the datasets considered in Sections \ref{ssec:cat}-\ref{ssec:power} (they range from just under 500 to just over 1500). This time the situation is quite the opposite to the one in Table \ref{tab:sim:power}. The 10\%-1\% range leads to accurate coverage rates for the Weibull law, while the rejection rate tends to be too high if the 25\%-2.5\% range is used (due to a poor fit in the tail of the distribution). For the power law type distributions, the percentage of observations identified as outliers significantly exceeds the expected rejection rates. Only the tails of the hyperbolic law, with their exponential decay, are hard to distinguish from the stretched exponential tails (note, that for $\tau=1$, the Weibull law reduces to the exponential distribution).

Summing up, the 10\%-1\% range of largest observations used to fit the power law or the stretched exponential tail seems to be a good choice. Only in the former case, the resulting CI can be overly conservative yielding a lower number of observations classified as outliers. Thus, once an observation is classified as a dragon king it is very likely to be so. On the other hand, if the power law fit is not accurate, an observation inside but close to the egde of the CI may be a dragon king as well.

\subsection{Catastrophe claims}
\label{ssec:cat}

\begin{figure}
\centering
\resizebox{0.9\columnwidth}{!}{\includegraphics{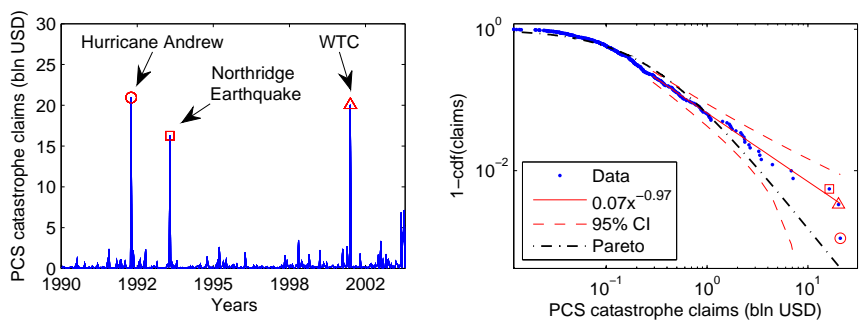}}
\caption{\emph{Left panel:} PCS catastrophe loss data, 1990-2004. The three largest losses in this period were caused by Hurricane Andrew (24 August 1992), the Northridge Earthquake (17 January 1994) and the terrorist attack on WTC (11 September 2001). \emph{Right panel:} Right tail of the empirical distribution of claim sizes. The three largest losses do not deviate significantly from the fitted power law. However, two out of three seem to be outliers with respect to the Pareto law fit.}
\label{fig:PCS}
\end{figure}

Severities of catastrophic events are known to exhibit a heavy-tailed behavior, even after excluding the most extreme outliers \cite{che:etal:06,hae:lop:10}. We will now check whether these outliers can be considered as dragon kings. To this end, we study the Property Claim Services (PCS) dataset, which covers losses resulting from catastrophic events in the U.S. The data include 1990-2004 market loss amounts in USD, adjusted for inflation using the monthly values of the Consumer Price Index (CPI). Only natural events -- except for the terrorist attack on World Trade Center, WTC, and the August 2003 blackout -- that caused damage exceeding five million dollars (nominal value, i.e.\ unadjusted for inflation) were taken into account. The three largest losses in this period were caused by Hurricane Andrew (24 August 1992), the Northridge Earthquake (17 January 1994) and the terrorist attack on WTC (11 September 2001), see the left panel in Figure \ref{fig:PCS}.

An earlier study \cite{bur:wer:08} of the PCS dataset spanning a five-year shorter time period (1990-1999) revealed that the two largest losses of the 1990s (Hurricane Andrew and the Northridge Earthquake) were outliers. At least such a conclusion could be drawn from the Pareto probability plot, see Figure 7.5 in \cite{bur:wer:08}. Using the technique developed in Section \ref{sec:test}, we want to check whether this assertion is justified. To this end, we plot the right tail of the empirical distribution of claim sizes, see the right panel in Figure \ref{fig:PCS}, and fit a power law to the 10\%-1\% largest observations. 

The estimated power law exponent of $p=-0.97$ indicates a very heavy tailed distribution (with a decay in the tail comparable to the Cauchy law). However, none of the three largest claim sizes exceed the 95\% CI allowing us to conclude that they are not dragon kings, `merely' black swans. 
Apparently, while the Pareto law exhibits a power law tail, the fit to the whole PCS dataset is far from perfect. Indeed, the reported maximum likelihood estimate of the Pareto law exponent for the 1990-1999 dataset was twice higher, i.e.\ the slope of the tail was much steeper \cite{bur:wer:08}. With such a fast decay, the two largest claims in the 1990s would seem to be outliers. However, performing the test introduced in Section \ref{sec:test} after fitting a power law only to the tail of the distribution (as in the right panel of Figure \ref{fig:PCS}) results in the rejection of the dragon king hypothesis.
 
\newpage

\subsection{Financial drawdowns}
\label{ssec:drawdowns}

Johansen and Sornette \cite{joh:sor:01} claim that dragon kings are common in the distributions of financial drawdowns. In this Section we test whether this is really the case.
Following \cite{joh:sor:01}, we define a drawdown as the loss between the local price maximum and the following local minimum. It is calculated as the percentage difference between the lowest price and the highest price in a decline period, i.e. the time series $1,2,5,4,3,3,1,3,4,3,2,3$ would result in two drawdowns, namely $\frac{1-5}{5}$ and $\frac{2-4}{4}$. 

The dataset analyzed here comprises 7661 NASDAQ index closing values from the period Feb. 5, 1971 -- May 31, 2000. The data were sampled at daily frequency and obtained from the Reuters EcoWin database. Although the time period studied seems to be the same as the one in \cite{joh:sor:01}, denoted there by [1971.1:2000.5], most likely it is not identical. The number of drawdowns we study is 1543, compared to 1495 in \cite{joh:sor:01}. Perhaps the precision of the index closing values is not the same (two decimal places in our dataset) or the ways of computing the drawdowns are slightly different.

\begin{figure}
\centering
\resizebox{0.9\columnwidth}{!}{\includegraphics{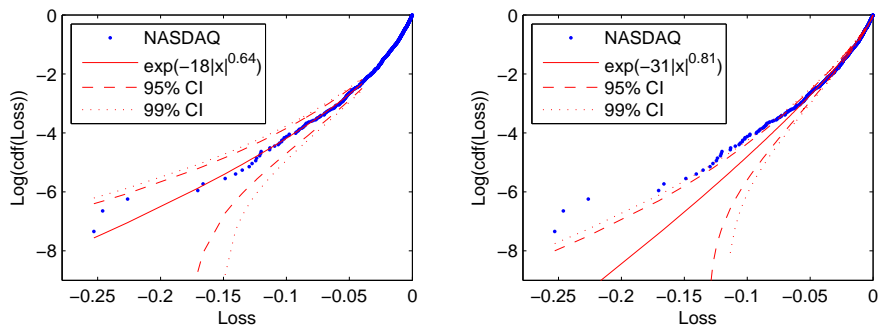}}
\caption{\emph{Left panel:} Stretched exponential (or Weibull) fit to the 10\%-1\% largest NASDAQ drawdowns from the period Feb. 5, 1971 -- May 31, 2000. Even the most extreme drawdowns cannot be classified as outliers. \emph{Right panel:} Weibull fit to all NASDAQ drawdowns from the same period. The estimated exponent of 0.81 is roughly the same as obtained in \protect\cite{joh:sor:01}. Nearly all of the largest 10\% of drawdowns deviate significantly from the Weibull law, indicating that perhaps a heavier-tailed distribution would yield a better fit.
}
\label{fig:crash}
\end{figure}

In the left panel of Figure \ref{fig:crash} we plot the stretched exponential (or Weibull) fit to the 10\%-1\% largest NASDAQ drawdowns. Apparently, even the most extreme drawdowns cannot be classified as outliers in this case. However, if we follow the same approach as Johansen and Sornette \cite{joh:sor:01} and using maximum likelihood calibrate the Weibull distribution to all NASDAQ drawdowns we obtain a nearly identical fit: $\beta\approx 31$ and $\tau\approx 0.81$; compare with the values in Table 2 in \cite{joh:sor:01}. This time almost all of the largest 10\% of drawdowns deviate significantly from the Weibull law, see the right panel in Fig. \ref{fig:crash} and compare it with Fig. 15 in \cite{sor:09}. In our opinion this is not evidence for dragon kings, but rather indicates that a heavier-tailed distribution would better describe the data. At the same time, this simple exercise shows that dragon kings are a model dependent feature. Depending on the model (stretched exponential distribution vs. tail) the same observations can or cannot be classified as outliers.

\subsection{Electricity spot prices}
\label{ssec:power}

\subsubsection{Why do they spike?}

\begin{figure}
\centering
\resizebox{0.9\columnwidth}{!}{\includegraphics{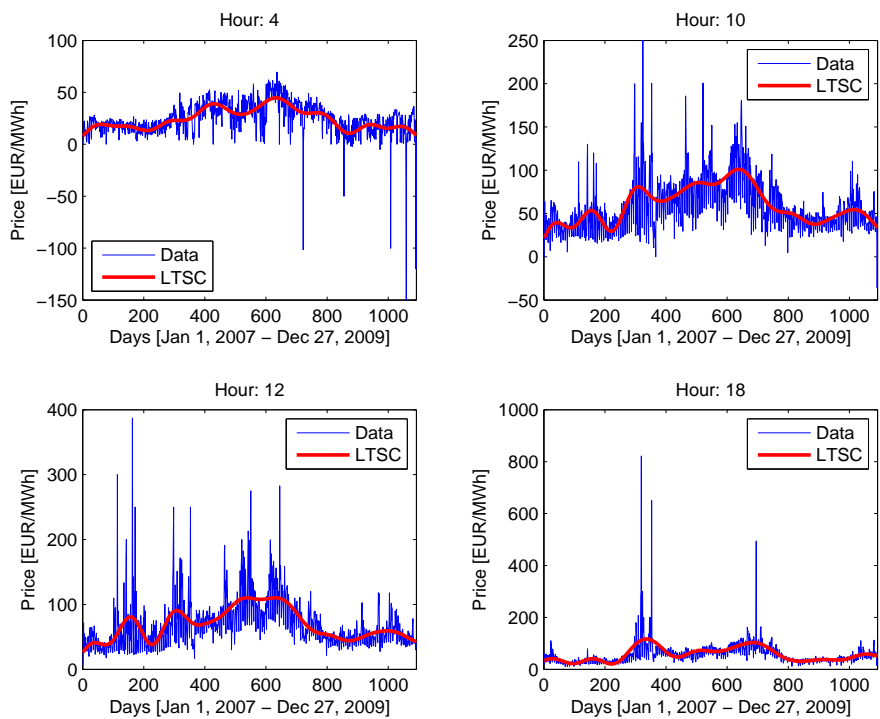}}
\caption{Electricity spot prices and the estimated long term seasonal components (LTSC) in the German EEX market for four sample hours and 1092 days in the period Jan. 1, 2007 -- Dec. 27, 2009. Note, the different y axis scales in the four panels.}
\label{fig:data:EEX}
\end{figure}

\begin{figure}
\centering
\resizebox{0.9\columnwidth}{!}{\includegraphics{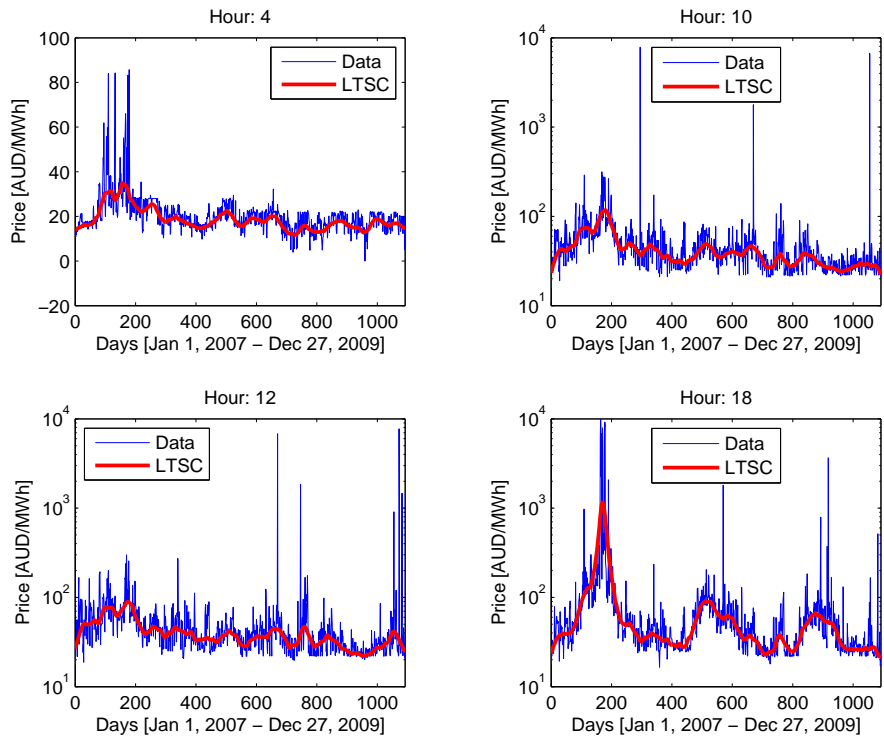}}
\caption{Electricity spot prices and the estimated LTSC in the Australian NSW market for four sample hours (in fact, half-hourly intervals starting at full hours) and 1092 days in the period Jan. 1, 2007 -- Dec. 27, 2009. Note, the semilogarithmic scale for hours 10, 12 and 18. The few extremely spiky prices would render the `normal' prices invisible on a linear scale.}
\label{fig:data:NSW}
\end{figure}

Electricity is a unique commodity and the power markets exhibit behavior like no other financial or commodity markets. Severe weather conditions, often in combination with the execution of market power by some players, have led in the recent past to spectacular price fluctuations -- ranging even two orders of magnitude within a matter of hours, see Figures \ref{fig:data:EEX} and \ref{fig:data:NSW}. These abrupt and short-lived price changes are known as spikes and are one of the most profound features of electricity spot prices \cite{bot:sap:sec:05,bys:05,wer:06}. 
But why do electricity spot prices spike in the first place?

The answer lies in the way these prices are determined. The spot price is typically set in a one- or two-sided uniform-price auction for each hour or a half-hourly time interval of the next day. It is determined as the intersection of the supply curve constructed from aggregated supply bids and the demand curve constructed from aggregated demand bids (for two-sided auctions) or system operator's demand forecast (for one-sided auctions). 
On the supply side of the market, all generation units of a utility or of a set of utilities in a given region are ranked (and form the so-called supply stack). This ranking is based on many factors, such as the marginal cost of production and the response time. The utility will typically first dispatch nuclear and hydro units, if available, followed by coal units. These types of plants are generally used to cover the so-called base load, whereas oil-, gas-fired and hydro-storage plants are used to meet peak demand. 

Demand, on the other hand, exhibits seasonal fluctuations reflecting variable business activities 
and changing climate conditions. In Northern Europe and Canada the demand peaks normally in the winter because of excessive heating. In 
Australia or mid-western U.S. it peaks in the summer months due to air-conditioning. Unexpected weather conditions can cause sudden shocks with demand typically falling back to its normal level as soon as the underlying weather phenomenon is over. 

The spot price is not very sensitive to demand shifts when the demand is low, since the supply stack is flat in the low-demand region. However, when demand is high and a larger fraction of power comes from `expensive' sources, even a small increase in consumption can force the prices to rise substantially. Then, when the demand drops, the price can rapidly decrease to the normal level. 
Likewise, if the consumption stays almost constant, price spikes can still appear when a considerable amount of `cheap' generation is withdrawn from the market (due to outages, maintenance, etc.).

While the supply-demand equilibrium explains price volatility it does not, however, justify the extreme severity of the spikes. It is not simply a matter of higher marginal costs. Rather, the spikes are a result of the bidding strategies of the market participants. Since electricity is a non-storable commodity, some agents are willing to pay almost any price to secure a sufficient and continuous supply. On a regular basis they place bids at the maximum level allowed \cite{wer:06}. The risk of having to pay the maximum price is relatively low, because in uniform-price auctions the spot price is what a buyer has to pay for each unit of power irrespective of what he or she did bid initially, as long as the bid was not less than the spot price.

\subsubsection{Dragon kings or black swans?}

We will now try to answer the question whether electricity price spikes should be classified as dragon kings or `only' as black swans. Or perhaps they are so common that they cannot be even regarded as black swans. To this end, we analyze the hourly (day-ahead) spot prices from two major power markets: the European Energy Exchange (EEX; Germany) and the New South Wales region of the National Electricity Market (NSW; Australia), see Figures \ref{fig:data:EEX} and \ref{fig:data:NSW}. For each market the sample totals 1096 daily observations for each hour of the day ($h=1, 2, ..., 24$) and covers the 3-year period Jan. 1, 2007 -- Dec. 27, 2009.  

It is well known that electricity spot prices exhibit several characteristic features, which have to be taken into account when analyzing or modeling such processes \cite{eyd:wol:03,wer:06}. These include seasonality on the annual, weekly and daily level. To cope with it we use the standard time series decomposition approach and let the electricity spot price $P_t$ for a particular hour $h$ be represented by a sum of two independent parts: a predictable (seasonal) component $f_t$ and a stochastic component $X_t$, i.e.\ $P_t = f_t + X_t$.  
Following \cite{wer:09:mmor} the deseasonalization is conducted in three steps. First, the long term seasonal component (LTSC) $T_t$ is estimated from the spot price $P_t$ using a wavelet filter-smoother of order 6 (for the EEX prices; for details see \cite{tru:wer:wol:07}) or a Gaussian kernel smoother with a bandwidth of $2^6$ (for the NSW prices). In the latter case the kernel smoother is used instead of the wavelet one due to it's lower sensitivity to extreme observations. 
A single non-parametric LTSC is used here to represent the long-term non-periodic fuel price levels, the changing climate/consumption conditions throughout the years and strategic bidding practices. As shown by Janczura and Weron \cite{jan:wer:10:ee}, the wavelet-estimated LTSC pretty well reflects the `average' fuel price level, understood as a combination of natural gas, crude oil and coal prices. 

\begin{figure}
\centering
\resizebox{0.9\columnwidth}{!}{\includegraphics{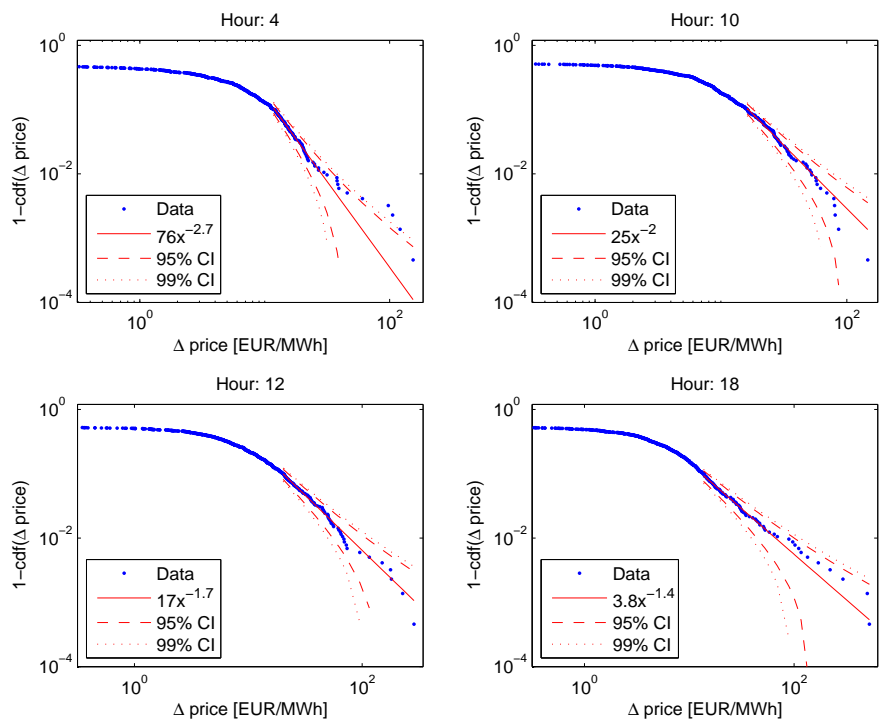}}
\caption{
Right tails of the empirical distribution of electricity spot price changes, computed for EEX prices depicted in Figure \protect\ref{fig:data:EEX}. The solid lines represent the fitted power law tails (to the lowest or highest 1\%-10\% of observations). The dashed and dotted curves indicate the 95\% and 99\% CI, respectively.}
\label{fig:tail:EEX}
\end{figure}

\begin{figure}
\centering
\resizebox{0.9\columnwidth}{!}{\includegraphics{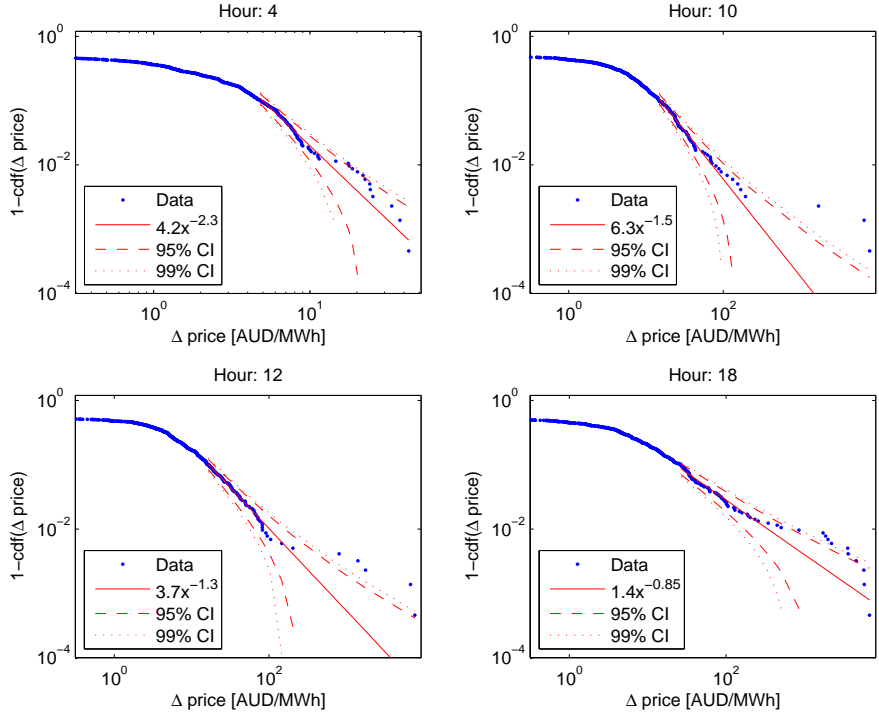}}
\caption{
Right tails of the empirical distribution of electricity spot price changes, computed for NSW prices depicted in Figure \protect\ref{fig:data:NSW}. The solid lines represent the fitted power law tails (to the lowest or highest 1\%-10\% of observations). The dashed and dotted curves indicate the 95\% and 99\% CI, respectively.}
\label{fig:tail:NSW}
\end{figure}

The price series without the LTSC is obtained by subtracting the $T_t$ approximation from $P_t$. Next, weekly periodicity $s_t$ is removed by subtracting the `average week' calculated as the arithmetic mean (for EEX data) or median (for NSW data; again for the sake of robustness) of prices corresponding to each day of the week. Additionally for the EEX dataset, the German national holidays are treated as the eight day of the week. Finally, the deseasonalized prices, i.e. $X_t = P_t - T_t - s_t$, are shifted so that the minimum of the new process $X_t$ is the same as the minimum of $P_t$. 

In Figures \ref{fig:tail:EEX} and \ref{fig:tail:NSW} we present the dragon king test results for the right tails of the distributions of deseasonalized price changes, i.e.\ of $\Delta X_t$, for four selected hours of the day. The more common returns (or changes of the log-prices) cannot be analyzed due to the existence of negative spot prices, which is another peculiarity of electricity markets. Two interesting conclusions can be drawn from these two figures. 

First, looking at the price trajectories in Figures \ref{fig:data:EEX}-\ref{fig:data:NSW} and the estimated power law exponents in Figures \ref{fig:tail:EEX}-\ref{fig:tail:NSW} it is apparent that Australian prices are more volatile and exhibit heavier-tails than their German counterparts. This can be explained by the fact that Australian markets operate as `energy only' markets, meaning that the wholesale electricity price should provide compensation to investors for both variable and fixed costs. Indeed it does. For instance, in South Australia the installed capacity increased by nearly 50\% in the period 1998-2003, almost half of it being open cycle gas turbines (OCGT) for peaking purposes \cite{wer:06}.

Second, in the EEX market only night hours (in particular 4 a.m.) yield price changes that can be classified as outliers, see the upper left panel in Figure \ref{fig:tail:EEX}. This is due to extreme negative prices on some days, possibly resulting from relatively large wind farm generation and very limited demand at this time of night. Somewhat surprisingly, the most spiky afternoon peak hours (like 6 p.m.) do not lead to dragon kings. The price changes on some days are extreme, but they do not deviate significantly from the fitted power law, see the lower right panel in Figure \ref{fig:tail:EEX}. 
On the other hand, the Australian NSW market is abundant in dragon kings. Nearly all daytime hours (including the three depicted in Figure \ref{fig:tail:NSW}) yield outliers, deviating significantly from the power law tails. Even some night hours (like 4 a.m.) exhibit extreme price changes that are nearly dragon kings.


\section{Conclusions}
\label{sec:conc}

In this short note we have developed a simple test for deviations from power law tails, which is based on the asymptotic properties of the empirical distribution function. We have used it to test whether great natural disasters, financial crashes or electricity price spikes should be classified as dragon kings or `only' as black swans. While not every observation deviating from a power law (or stretched exponential) fit can be called a dragon king, the bottom line is that outliers to power law tails exist. Following Sornette \cite{sor:09} we can call them `dragon kings'.

Interestingly, Sornette actually goes one step further and argues that dragon kings may have properties that make them predictable. Now, this is controversial. It's one thing to identify dragon kings but quite another to spot the event that triggers a crash or a price spike. The assertion that dragon kings are more easily predictable than other events requires proofs, which go well beyond the scope of this short note. 

Let us just mention one fact that might justify this line of thought. Recently Cartea et al. \cite{car:fig:gem:09} have shown in the context of power market modeling that predicting the timing of electricity price spikes is possible, at least to some extent, when forward looking information on capacity constraints is taken into account. Unfortunately, in most power markets the availability (to every power market participant) of the reserve margin data is limited. 

\section*{Acknowledgments}
J.J. acknowledges partial financial support from the European Union within the European Social Fund. The work of R.W. was partially supported by ARC grant no. DP1096326

\end{document}